# RePaCA: Leveraging Reasoning Large Language Models for Static Automated Patch Correctness Assessment

**Marcos Fuster-Pena**[1], **David de-Fitero-Dominguez**[1], **Antonio Garcia-Cabot**[1,*], **Eva Garcia-Lopez**[1]

[1]*Departamento de Ciencias de la Computación, Universidad de Alcalá, Alcalá de Henares, Madrid, Spain*

**Corresponding author: a.garciac@uah.es*

## ABSTRACT

Automated Program Repair (APR) seeks to automatically correct software bugs without requiring human intervention. However, existing tools tend to generate patches that satisfy test cases without fixing the underlying bug. These are known as overfitting patches. To address this issue, Automated Patch Correctness Assessment (APCA) attempts to identify overfitting patches generated by APR tools. This problem can be solved using a static approach, meaning that no additional information is needed beyond the original and fixed code snippets. Current static techniques often struggle with reliability, flexibility and transparency. To address these issues, we introduce RePaCA, a reasoning-based static APCA technique that leverages Large Language Models (LLMs) specialized in thinking tasks. Our model is prompted with both buggy and fixed code snippets and guided to generate a Chain of Thought that analyzes code differences, reasons about how the patch addresses the root cause, and ultimately provides a binary classification: correct or overfitting. To enhance these reasoning capabilities for the APCA task specifically, the LLM is fine-tuned using Reinforcement Learning with the Group Relative Policy Optimization algorithm. When evaluated on a standard Defects4J-derived test, our approach achieves state-of-the-art performance, with 83.1% accuracy and an 84.8% F1-score. Furthermore, our model demonstrates robustness to domain shift when trained on different datasets, outperforming the leading technique. This reasoning capability also provides explicit rationales that can support the inspection of patch assessments. These findings underscore the considerable promise of fine-tuned, reasoning LLMs to advance static APCA by enhancing accuracy and transparency.

## 1 INTRODUCTION

Automated Program Repair (APR) consists of identifying and fixing program bugs automatically, reducing human debugging effort in software development and maintenance. Modern APR tools often use deep learning to learn bug-fixing patterns from code repositories, treating repair tasks as translation problems where buggy code is converted into fixed versions [1].

Typically, these tools first locate the suspicious fragments where a bug can be found, such as buggy methods or functions. They then modify the code and generate one or more patches to fix the bug. Finally, they verify that these changes are correct, usually using test cases. Candidates that pass all the tests are called *plausible patches* [2]. However, passing all the tests does not mean that the patch fixes the bug, since test cases may not always ensure that a program works correctly. To intuitively understand this, for example, if a tool detects that the functionality of a program is generating a bug, the tool can generate a patch that eliminates the functionality, fixing the bug but affecting the overall operation of the program. Thus, patches that pass the tests and fix the bug are called *correct patches*; otherwise, they are called *overfitting patches* [3].

Therefore, one of the most critical challenges in APR is identifying the *overfitting patches* generated by the tools [4]. Consequently, a variety of complementary tools have been proposed to solve this problem automatically. This subfield of APR is called Automated Patch Correctness Assessment (APCA). This process is useful because it acts as a filtering step after patch generation, reducing the number of incorrect plausible patches that must be manually inspected and lowering the risk of accepting patches that pass the available tests but do not correctly fix the bug. These techniques can be classified, based on how they address the problem, into two categories [3]:

- Dynamic approach: this approach consists of analyzing the runtime information provided by the patch when executed on the tests to determine the correctness of a particular patch [5]. In addition, automated test generation tools can be used to assess patches [6], [7]. This means that, before determining whether a patch fixes a bug or not, the patch must be executed and tested, which is why these approaches are considered complex and time-consuming [5], [8].
- Static approach: in this case no information about the execution of the patch is provided, focusing on the comparison of the code with bugs (buggy code) and the code with the patch that tries to fix the bug (fixed code). In general, this approach tries to analyze the characteristics and differences between the two versions of the code, allowing it to recognize patterns of overfitting. This technique has proven to be the simplest but tends to be less accurate due to the limited amount of information available [9], [10], [11], [12], [13].

Having reviewed both dynamic and static approaches, this article focuses exclusively on static techniques, those that do not rely on execution information, which is generally more expensive to obtain than the patch code [14], [15].

Our proposal consists of implementing a Large Language Model (LLM) specialized in reasoning to evaluate code patches. We consider LLMs particularly suitable for static APCA because they can process source code as structured natural-language-like input, compare buggy and patched snippets, and reason about whether a change addresses the underlying bug or only matches specific cases [14]. Our methodology provides the model with both *buggy* and *fixed code snippets,* guided by a specific prompt. This prompt encourages the generation of a Chain of Thought (CoT), which is an internal step-by-step monologue that mimics human introspective reasoning, that analyzes the input step by step, examining the differences between the code versions, reasoning about the features of the patch and concluding with a binary answer. This approach is based on recent studies showing that extended inference in LLMs can significantly improve the quality of model answers [16].

To enhance these reasoning capabilities, specifically for the APCA task, we employ Reinforcement Learning (RL). Instead of relying on a single absolute sample, we evaluate multiple possible solutions simultaneously and reward the model based on its relative performance. This is achieved by utilizing the Group Relative Policy Optimization (GRPO) algorithm for fine-tuning [17]. This approach has achieved significant results in other areas, such as mathematics and code generation [17], [18]. This combination adapts recent reasoning-oriented LLM training techniques to the APCA domain, where their applicability and effectiveness remain underexplored. In this context, our work evaluates whether GRPO-trained reasoning LLMs can improve static patch correctness assessment.

We hypothesize that most static APCA tools cannot understand the root cause of complex overfitting patches, so deep reasoning capabilities of LLMs can increase effectiveness in this field. If the majority of overfitting problems presented can be solved via a reasoning process, the LLM will analyze the differences between both code snippets, consider all edge cases and potential issues, and assess how the proposed fix addresses the root cause. It will also evaluate whether the solution is broadly applicable or only works for specific inputs, providing a thorough explanation of its reasoning.

Our experiments show that our reasoning-based LLM outperforms other static APCA techniques across the reported metrics. This improvement comes from the LLM's understanding of the actual effect of a code change and the nature of the overfitting problem, whereas other techniques suffer from a more superficial analysis. Furthermore, when compared to the leading state-of-the-art static APCA technique, our model shows stronger cross-dataset performance under the evaluated setting, suggesting that reasoning-oriented LLMs may adapt better to patches from different sources. In contrast to most existing APCA techniques, RePaCA also generates a human-readable rationale through its <*think*> block. This rationale can support manual inspection and help developers analyze the prediction, although it should not be interpreted as a faithful explanation of the internal decision process that produced it.

In summary, identifying overfitting patches remains a critical challenge in Automated Program Repair. To address the limitations of current static APCA techniques, this study introduces RePaCA, which leverages reasoning-oriented LLMs fine-tuned via the GRPO algorithm. This approach enables a deep, step-by-step analysis of code differences, aiming to significantly enhance both assessment accuracy and transparency.

# 2 RELATED WORK

Static APCA tools focus on analyzing code patches without needing runtime information, relying only on the original and patched code. Several approaches have been developed in this area. For instance, ODS [12] primarily extracts static code features at the Abstract Syntax Tree (AST) level, which is a hierarchical representation that acts as a structural map of the code's grammar, and employs traditional supervised learning for classification, focusing on identifying patterns in code changes indicative of overfitting. Similarly, CACHE [9] emphasizes learning context-sensitive embeddings of code changes by considering program structure, particularly through AST paths, to capture more semantic information by observing changes in their surrounding context.

Additionally, Quatrain [13] offers a unique perspective by framing patch assessment as a question-and-answer problem. It utilizes Natural Language Processing (NLP) techniques to evaluate the semantic relationship between a bug report (viewed as the question) and a description of the patch (considered as the answer).

Neural-network models have been applied across diverse computational settings, including dynamic control and state-estimation problems [19], [20]. Within software engineering, LLMs have made significant strides in recent years, demonstrating their effectiveness across multiple domains. Their massive size and training on vast text corpora allow them to capture linguistic patterns and knowledge on a large scale. This ability to understand text has opened the door to new applications, such as APCA. For example, INVALIDATOR [10] takes a different approach by combining program semantics based on program invariants (inferred from test sets) with syntactic processing using pretrained language models like CODEBERT [21]. This dual approach allows it to verify both behavioral and syntactic correctness.

More recently, APPT [11] has advanced the field by leveraging and fine-tuning large pre-trained language models such as BERT [22]. APPT treats code as a sequence of tokens, learns task-specific representations for assessing patch correctness, applies a Long Short-Term Memory (LSTM) layer to capture sequential dependencies, and finally uses a classification layer for its decision. This technique demonstrated the highest performance among static APCA methods prior to the work described in this article. Furthermore, other recent studies have explored the use of Large Language Models (LLMs) for code through in-context learning techniques. A representative approach in this direction is LLM4PatchCorrect [14], which infers patch correctness without the need for fine-tuning by concatenating bug information with semantically similar labeled patches to form the input demonstration.

These techniques can be grouped into methodological families according to the type of evidence they exploit: statically extracted code features [12], learned embeddings of code changes and their context [9], program invariants paired with syntactic encoders [10], the correlation between a bug report and a textual patch description [13], or pretrained language models, whether fine-tuned on token sequences [11] or conditioned in context on retrieved labeled patches [14]. Despite these differences, all of them classify a patch by matching syntactic or structural regularities and learned correlations, rather than by explicitly reasoning about whether the change resolves the root cause of the bug. This shared limitation motivates our approach, which frames patch correctness as a multi-step reasoning process over the buggy and fixed snippets instead of a single-pass pattern-matching decision.

This kind of multi-step reasoning has become feasible thanks to modern LLMs, which have shown extraordinary performance in solving complex problems through Test-Time Compute (TTC) and multi-stage thinking. Models like DeepSeek R1 [18] integrate the CoT prompting into their architecture to generate intermediate reasoning steps, thereby enhancing accuracy in tasks such as mathematics and code generation. These models are further trained through RL. For example, Proximal Policy Optimization (PPO) [18], a policy gradient method, has gained popularity due to its balance between performance and computational efficiency. More recently, GRPO [17] has been shown to further improve the accuracy of LLMs in reasoning tasks. GRPO is an advanced policy gradient algorithm that builds on the principles of PPO. GRPO optimizes policies by constraining updates to ensure stability while improving group sample efficiency. These developments motivate our evaluation of reasoning-oriented RL fine-tuning within the APCA setting.

In summary, while existing static APCA tools have successfully integrated large language models to analyze semantic and syntactic code features, they have yet to fully exploit multi-stage reasoning capabilities. By adapting advanced reinforcement learning techniques like GRPO for explicit Chain of Thought generation, our work addresses this gap in the APCA domain.

# 3 METHODOLOGY

This section outlines the systematic approach used to guide our model through the entire automated patch assessment pipeline. First, we explain the prompting strategy used to guide the model in assessing each patch; subsequently, we describe the reinforcement learning training loop that rewards both the form and accuracy of the outputs. Additionally, we describe the selection process for the base model size. Finally, we detail the computational setup and hyperparameter choices to ensure the reproducibility of our experiments.

## 3.1 PROMPT DESIGN

To improve the quality of the model's answers, it is essential to design a clear, structured and specific prompt that adequately guides the reasoning process. In the context of our task, the design of the prompt is particularly critical. This requirement becomes even more significant when working with smaller models, where explicit guidance in the prompt can significantly impact performance [16]. The model's input as well as the output is represented in Figure 1.

The prompt begins with a brief explanation of the problem, describing the existence of two code snippets (the buggy version and the proposed fix) and defining the model's objective: to identify whether the changes truly solve the underlying problem or merely adjust the behavior to pass specific test cases, resulting in an *overfitting patch*. This is followed by the analysis requirements, which guide the model on how to approach the evaluation. These steps are intended to encourage deep and structured reasoning.

Another important aspect of the prompt is the specification of the output format, which is the same as that used in DeepSeek-R1 [18], and allows structuring the model response in two parts: an internal monologue, within the labels <think>...</think>, where the reasoning is developed step by step, and a final conclusion, within <answer>...</answer>. This design leverages the CoT approach mentioned earlier to mimic an introspective thought process before making a final judgment, which helps to better align the responses with the defined evaluation criteria. The *think block* is particularly important, as its analysis will provide insights into the model's ability to perform logical steps aligned with the desired analytical process. Nevertheless, it is important to consider that our model may not provide an answer or may provide an answer in the wrong format, although this is a minor problem because only around 1% of the given prompts were answered incorrectly through the testing process on the small dataset. To address this fault in a real-world environment, re-sampling can be a simple and effective option. Additionally, we implemented the one-shot prompting technique, providing the model with a solved example in the same required format. This approach allows the model to more clearly identify the procedure to follow, improving the consistency of answers. The effectiveness of this technique has been extensively documented in previous works on one-shot learning and example-based reasoning [18].

Finally, the sample data (*buggy code* and *fixed code*) to be analyzed are provided to the model. This prompt concludes by instructing the model to begin its reasoning, thereby triggering the process described above.

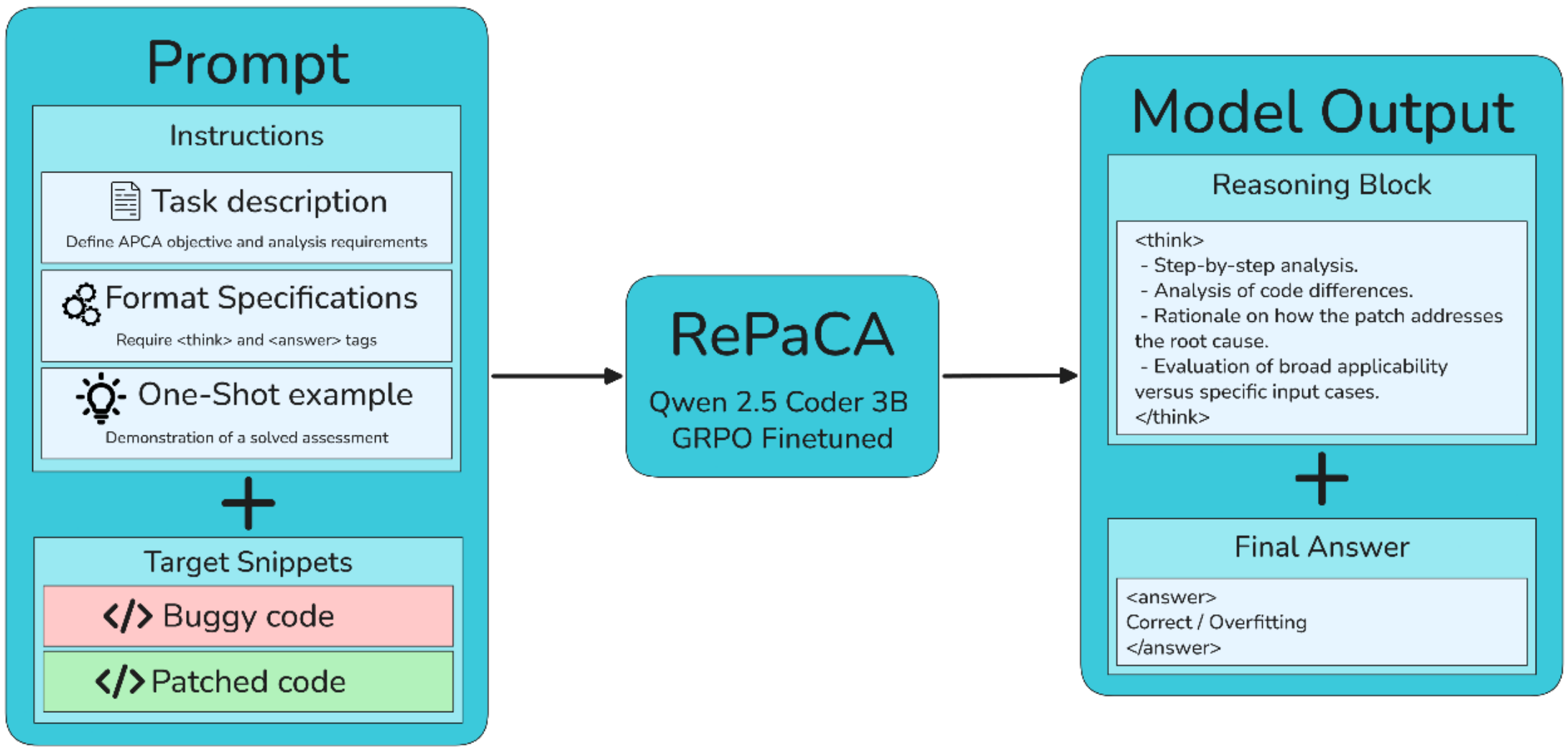

*Figure 1: Overall structure of the prompt provided to the model, alongside the expected output format detailing the reasoning and final classification.*

## 3.2 Training Architecture

This section describes our reinforcement-learning training pipeline using GRPO. We explain the algorithmic framework of this approach. Then, we detail the reward functions used and how we score each generated output, using structural and accuracy-based metrics to guide the model toward coherent reasoning traces and correct answers.

### 3.2.1 Reinforcement Learning Policy

To test the viability of RL-only adaptation in this domain, our approach optimizes a base LLM for APCA, without requiring any prior supervised fine-tuning. Unlike conventional supervised learning, which trains the model to imitate a fixed target output, RL optimizes generated responses according to task-specific rewards. This is advantageous in APCA because several reasoning paths may be valid for the same patch, while the desired behavior is that the model follows a structured format and reaches the correct final classification. Specifically, the GRPO algorithm [17] is employed, which is a variant of PPO that eliminates the need for a value model by leveraging the average reward of multiple sampled outputs. By comparing several candidate responses for the same prompt, GRPO provides a direct mechanism for aligning the model with the expected APCA behavior, namely structured reasoning followed by an accurate correctness assessment. This renders the approach more computationally efficient and stable, making it especially suited for generating CoT explanations. This training method is based on DeepSeek R1 Zero [17].

Figure 2 illustrates the fine-tuning pipeline utilizing GRPO. The framework uses a frozen reference model to regularize policy updates via a KL penalty, preventing model collapse. Optimization is driven by both rewards, format and accuracy rewards. By normalizing the cumulative rewards across a sampled group ($G = 8$), GRPO computes relative advantages ($A_x$) that provide gradient updates ($\nabla\theta$), effectively refining the model's reasoning for a better performance on the APCA task.

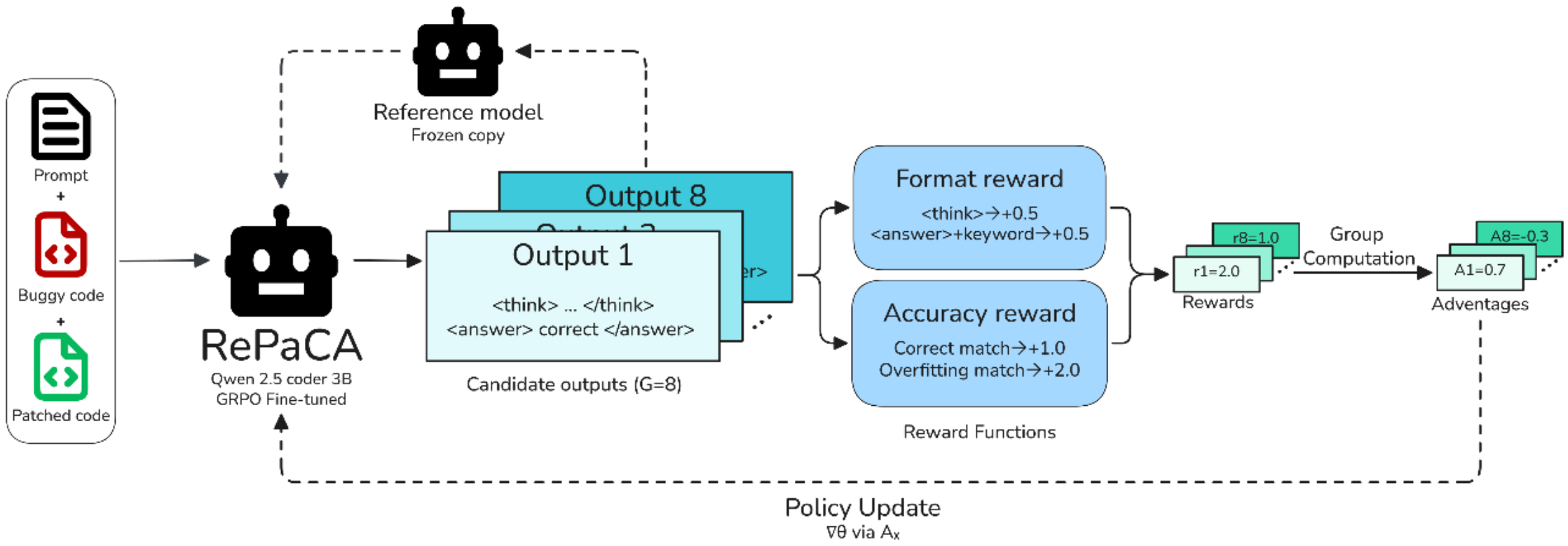

*Figure 2: GRPO training architecture*

### 3.2.2 Reward functions

To align the model’s behavior with a specific task, GRPO uses reward functions, which indicate the correctness of the model’s responses during the training process. To design the reward, our study has been inspired by previous approaches, such as those used in DeepSeek R1[18]. Our approach adopted two types of rewards: a format reward and an accuracy reward, which play complementary roles in guiding the model through the learning process.

The **format reward** aims to incentivize the model to generate structured outputs in a consistent and recognizable way. This reward is defined as a value between 0.0 and 1.0 and consists of three parts. First, if the prediction correctly contains the label <think>...</think>, 0.5 points are added to the reward. To this, 0.25 additional points are added if the output also includes a response encapsulated in the <answer>...</answer> tag. Finally, another 0.25 points are awarded if at least one keyword from among the following is found within that response: *“overfitting”, “fixed”, “fix”, “correct”, “buggy”, “incorrect”* or *“wrong”.*

It should be noted that this list of keywords was intentionally extended beyond the labels “correct” and “wrong” used in previous works [14], in order to provide more flexibility to the model in expressing its judgment. In this context, terms such as fix and fixed are considered equivalent to correct; just as buggy, incorrect or overfitting express essentially the same type of patch.

The **accuracy reward** focuses on measuring whether the classification made by the model regarding the patch type (correct or overfitting) matches the reference label. This reward can take values between 0.0 and 2.0, depending on the final answer of the model and the expected answer. As shown in Table 1, if the model prediction does not match the actual label, a reward of 0.0 is assigned. If it matches and the patch is correct, the reward is 1.0. Finally, if it matches and the patch is *overfitting*, the reward is set to 2.0.

This unequal reward design responds to the empirical observation that the model tends to have more difficulty in detecting overfitting patches than in identifying correct ones. The specific values for the accuracy reward (1.0 for correct, 2.0 for overfitting) were chosen after preliminary experimentation. We started with equal rewards, gradually increasing the difference, and observed the model’s learning curve and output correctness. The 2.0 value for correctly identifying an *overfitting patch* provided a sufficient incentive to improve performance in this harder class without introducing bias, thus maintaining a reasonable balance, as confirmed by our precision and recall metrics (see Section 4.1).

| Model Answer | Expected Answer | Reward |
|---|---|---|
| Overfitting | Correct | 0.0 |
| Correct | Overfitting | 0.0 |
| Correct | Correct | 1.0 |
| Overfitting | Overfitting | 2.0 |

*Table 1: Accuracy rewards between the model answer and the expected answer*

## 3.3 EXPERIMENTAL SETUP

### 3.3.1 Model selection

To balance predictive performance with computational feasibility, we constrained candidate backbones to lightweight instruct models (approximately 2B-3B parameters), targeting both stable GRPO training and practical deployment. In this setting, we prioritized class-balanced metrics over raw accuracy, since APCA requires reliably detecting overfitting patches without severely degrading precision. Therefore, the final backbone choice was based on the controlled ablation reported in Section 4.5.1. Following that analysis, we selected Qwen2.5-Coder-3B as the base model because it provided the strongest F1-score and AUC within our computing budget.

### 3.3.2 Evaluation metrics

To evaluate the model's performance, a binary classification is used (Correct = Negative, Overfitting = Positive, as shown in Table 2). We employ the most widely used metrics in APCA research: accuracy, precision, recall, F1-score and AUC [11]. These metrics leverage true positives, false positives, false negatives and true negatives to measure, respectively, the overall proportion of correctly classified patches; the reliability of labeling a patch as overfitting; the effectiveness in detecting all overfitting patches; the harmonic mean of precision and recall; and the classifier's ability to distinguish overfitting from correct patches across all decision thresholds. We selected these five metrics because they are standard in the literature, facilitate direct comparison between techniques, and capture both global accuracy and the balance between sensitivity and specificity.

| | Overfitting patch (Positive) | Correct patch (Negative) |
|---|---|---|
| **Predicted as Overfitting patch** | True positive (TP) | False positive (FP) |
| **Predicted as Correct patch** | False Negative (FN) | True Negative (TN) |

*Table 2 Patch classification confusion matrix*

#### 3.3.3 Datasets

The datasets used in this study are derived from CACHE [9] and were collected using APPT [11]. These datasets were extracted from files provided on GitHub[1], following the methodology outlined in the original paper. As shown in Table 3, the samples are divided into two datasets:

| Dataset | Correct | Overfitting | Total |
|---|---|---|---|
| **Small** | 535 | 648 | 1,183 |
| **Large** | 25,589 | 24,105 | 49,694 |

*Table 3 Datasets used from APPT [11][14].*

**Small:** This dataset contains 1,183 patches and was created by combining patches from two previous studies: one by Tian et al. [23] (1,000 patches) and another by Wang et al. [3] (902 patches). After filtering and removing duplicates, a total of 1,183 patches were obtained from the widely used Defects4J v1.2 [24] benchmark. This set focuses on Java-developed patches, as most automatic repair tools (APR) have been evaluated on Defects4J.

**Large:** This dataset contains 49,694 patches. Most correct patches come from ManySStuBs4J [25], which extracts correct patches from more than 1,000 open-source projects. Notably, these patches are sourced from open-source GitHub projects rather than APR tools. Conversely, overfitting patches were obtained from the RepairThemAll framework [26], which generates patches for a variety of bug benchmarks. Duplicated and biased patches were removed from both sources.

Both datasets are balanced and modest in size, particularly the small dataset. In addition, the standard 5-fold cross-validation method [9] is used to test the performance of different APCA techniques. This method, which involves dividing the dataset into five partitions, is widely used in APCA due to the small size of the datasets.

Additionally, to conduct a direct comparison with INVALIDATOR [10], we utilized its specific benchmark dataset comprising 885 patches. Following their evaluation protocol, this dataset is divided into a training set of 746 patches and a dedicated test set of 139 patches, using data available on their original codebase [2].

#### 3.3.4 Implementation Details

Our model was trained on a server with an NVIDIA H100 NVL GPU. However, after training, RePaCA uses a lightweight 3B-parameter LLM that can be run for inference on consumer GPUs with as little as 8GB of VRAM. We used the TRL library for GRPO training, as well as Unsloth[3] for training optimization. All experiments were implemented using the PyTorch framework.

Regarding the hyperparameters used, we trained for 20 epochs. This duration was determined to be generally optimal based on observing the convergence of the accuracy metric and the reward signals on a validation split of the training data during our development phase. Training beyond this point did not yield significant improvements and sometimes showed early signs of overfitting to the training set dynamics. This contrasts with APPT, which uses 50 epochs, potentially due to differences in model architecture, training algorithms, or dataset characteristics. To prevent overfitting, we used a learning rate of 5e-5 —matching that of APPT— which produced strong results. Similar learning rates have been employed in other GRPO training studies [27]. We set the batch size to eight samples, used two gradient-accumulation steps, and configured the GRPO policy to generate eight candidates. These choices were made to reduce VRAM consumption and training time without sacrificing training quality. For generation, as shown in Section 4.5.2, the temperature was set to 0.6 to balance creativity and determinism, encouraging diverse yet coherent and factually grounded responses while avoiding the repetitiveness of lower temperatures. Finally, the model's maximum completion length was set to 2,048 tokens, providing a sufficient context window that allows the model to reason effectively and generate accurate responses. To ensure a faithful baseline comparison, the reproduction of APPT's results was conducted using its original official codebase retrieved from the authors' public GitHub[4] repository. The hyperparameters used were those reported to yield the best performance in the manuscript.

[1] https://github.com/iSEngLab/APPT

[2] https://github.com/thanhlecongg/Invalidator

[3] https://unsloth.ai/

Overall, our methodology relies on a carefully designed prompt that elicits structured reasoning from a lightweight base model (Qwen2.5-Coder-3B). By exclusively utilizing the GRPO reinforcement learning algorithm with tailored format and accuracy rewards, the model is fine-tuned to accurately assess patch correctness using standard, balanced benchmark datasets.

# 4 EXPERIMENTAL RESULTS

In this section, we evaluate the performance of our model compared to the representative static APCA techniques in different scenarios:

1. Standard Evaluation: This initial experiment is designed to rigorously evaluate RePaCA's performance by following the standard 5-fold cross-validation protocol on the 'Small' dataset, the established benchmark in the APCA community. This approach ensures a direct and fair comparison against leading techniques like APPT and CACHE, which use the same evaluation framework. Furthermore, given the dataset's modest size, cross-validation is essential to guarantee robust and reliable results, mitigating the risk that the observed performance is due to a single favorable data split. Therefore, this experiment validates RePaCA's effectiveness and establishes its baseline performance in the most accepted scenario in the literature. To enable a comparison with INVALIDATOR, an additional evaluation was conducted using their original dataset.
2. Cross-Dataset Evaluation: We also present a performance comparison between our model and the highest-performing baseline (APPT) to study robustness to domain shift. Unlike CACHE, which trains on the large dataset and tests on the small dataset, we adopted this setup to simulate a realistic transfer scenario where high-quality curated data is scarce for training and the model has to adapt to other domains. Additionally, the large dataset has lower quality data because its correct patches are sourced from GitHub projects rather than APR tools. This can introduce variability or patterns distinct from APR tools, potentially leading to weaker training results and poorer performance on APR-generated patches. This benchmark demonstrates the model's robustness to domain shift.

[4] https://github.com/iSEngLab/APPT

3. Qualitative Analysis: Finally, we provide two examples of our model's behavior when classifying patches —one correct and one overfitting — both of which are correctly identified. We present the model's reasoning and explain why such reasoning is correct.

## 4.1 PERFORMANCE ON SMALL DATASET

The performance of the static APCA techniques evaluated under the same input setting for the small dataset is presented in Table 4. The models are listed in the first column, ordered by accuracy from lowest to highest. The subsequent columns report the evaluation metrics: accuracy, precision, recall, F1 and AUC, respectively. The metrics for CACHE and APPT were extracted from their original papers, while those for ODS and Quatrain were obtained from LLM4PatchCorrect [14], where precision and recall were not specified.

| | Accuracy | Precision | Recall | F1 | AUC |
|---|---|---|---|---|---|
| **ODS [12]** | 62.3 | - | - | 69.1 | 62.5 |
| **Quatrain [13]** | 69.4 | - | - | 77.8 | 64.5 |
| **CACHE [9]** | 75.4 | 79.5 | 76.5 | 78.0 | 80.3 |
| **APPT [11]** | 79.7 | 80.8 | 83.2 | 81.8 | 82.5 |
| **Ours** | **83.1** | **84.0** | **85.7** | **84.8** | **82.8** |

*Table 4 Performance of our model compared with static APCA techniques on the small dataset (best values for each metric are in bold)*

Our model demonstrates a significant improvement over existing techniques, achieving the best performance among the evaluated static APCA baselines. Notably, our model achieves an accuracy of 83.1%, surpassing APPT by 3.4 percentage points and CACHE by 7.8 percentage points. In terms of precision, our model scores 84.0%, outperforming APPT (80.8%) and CACHE (79.5%) by 3.2 and 4.5 points, respectively. Our recall of 85.7% also exceeds that of APPT (83.2%) by 2.5 points and CACHE (76.5%) by a substantial 9.2 points. Consequently, the F1-score of our model is 84.80%, exceeding APPT (81.8%) by 3.0 points and CACHE (78.0%) by 6.8 points. Regarding AUC, our model achieves 82.8%, slightly outperforming APPT (82.5%) and surpassing CACHE (80.3%).

These results highlight that our LLM surpasses the performance of previously leading static APCA techniques, such as APPT and CACHE, on this dataset. This underscores the effectiveness of our approach in accurately identifying both correct and overfitting patches, offering robust and well-balanced performance, as evidenced by the precision and recall metrics.

To further assess the statistical reliability of RePaCA's improvement under a paired setting, we reproduced the leading technique (APPT) on the small dataset. We performed this experiment using its official implementation and the same predefined 5-fold splits, since instance-level predictions from the original paper are not available. The APPT values in Table 5 therefore correspond to our reproduction rather than to the published aggregate metrics reported in Table 4. We then performed paired statistical tests between APPT and RePaCA. McNemar's test was computed on 1,163 paired instances. The remaining 20 instances (1.7% of the dataset) were excluded because RePaCA produced an unparseable output, meaning no binary prediction was available for the paired test. McNemar's test showed a statistically significant difference in favor of RePaCA ($\chi^2 = 25.90$, $p = 3.6 \times 10^{-7}$). To verify that the excluded cases did not affect the conclusion, we repeated the analysis by assigning these unparseable outputs to the majority class. The result remained statistically significant ($\chi^2 = 24.82$, $p = 6.3 \times 10^{-7}$). In addition, we computed bootstrap confidence intervals with 10,000 resamples for the metric differences. As shown in Table 5, all confidence intervals exclude zero, indicating statistically significant improvements.

| Metric | APPT (95% CI) | RePaCA (95% CI) | Diff (95% CI) |
|---|---|---|---|
| **Accuracy** | 75.0 [72.4, 77.5] | 83.1 [80.9, 85.2] | +8.08 [+5.1, +11.1] |
| **Precision** | 76.2 [72.9, 79.4] | 83.9 [81.1, 86.7] | +7.75 [+4.6, +11.0] |
| **Recall** | 79.4 [76.3, 82.5] | 85.7 [82.8, 88.3] | +6.24 [+2.3, +10.1] |
| **F1-score** | 77.8 [75.2, 80.2] | 84.8 [82.6, 86.9] | +7.02 [+4.3, +9.8] |

*Table 5 Bootstrap confidence intervals for APPT and RePaCA on the small dataset. Values are percentages. The confidence intervals were computed with 10,000 bootstrap resamples.*

## 4.2 PERFORMANCE AGAINST INVALIDATOR

To provide a direct comparison with INVALIDATOR [10], we evaluated RePaCA using their dataset and split configuration (746 training patches and 139 test patches). The results were compared against those reported in their original manuscript. Figure 3 presents the confusion matrices for both models on the test set.

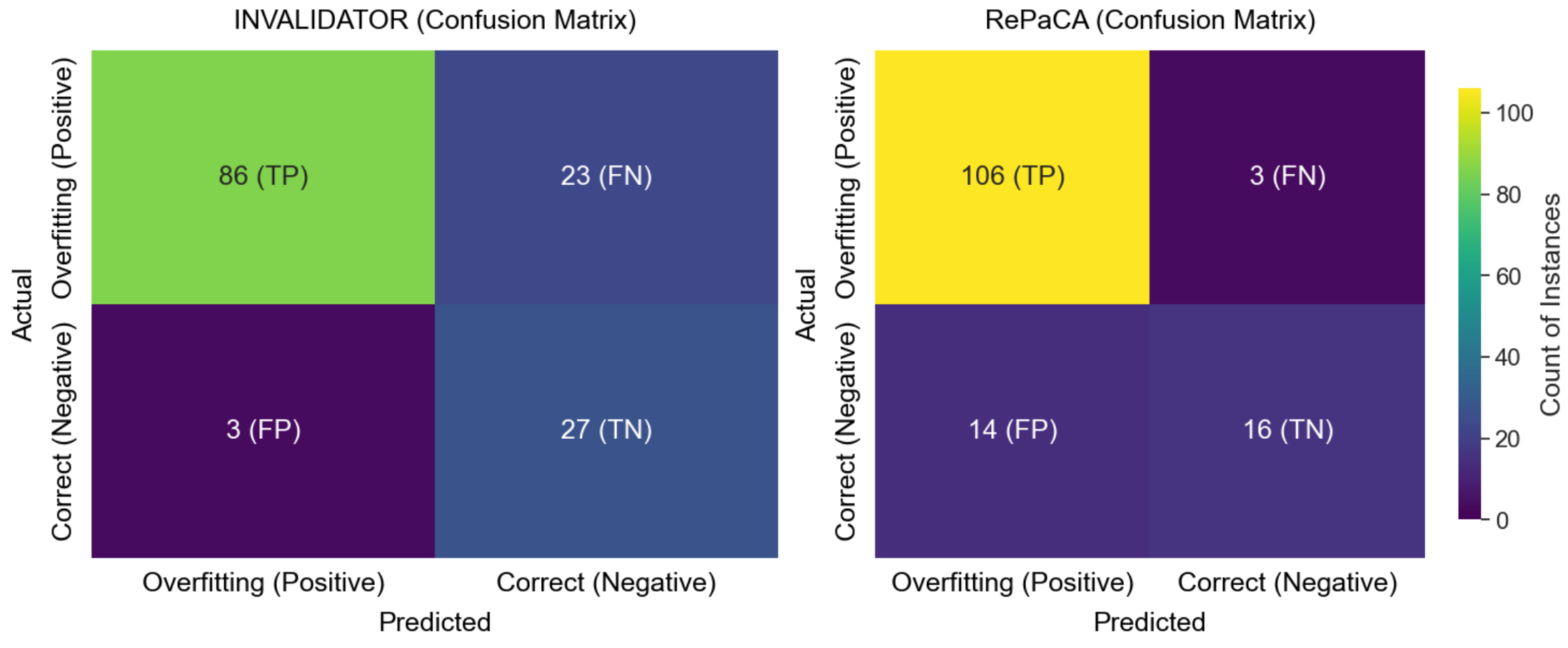

*Figure 3 Comparison of confusion matrices between INVALIDATOR and RePaCA on the 139-patch evaluation split.*

Comparing the performance metric by metric, our approach achieves an accuracy of 87.8%, outperforming INVALIDATOR's 81.2%. Regarding recall, RePaCA correctly identifies 106 out of 109 overfitting patches, reaching 97.2%, compared to INVALIDATOR's 78.9% (86 out of 109). Conversely, INVALIDATOR yields a higher precision (96.6% versus RePaCA's 88.3%), as it produces fewer false positives (3 compared to RePaCA's 14) within the correctly identified negative samples. Ultimately, our approach achieves a higher F1-score of 92.6%, surpassing INVALIDATOR's 87.0%.

## 4.3 PERFORMANCE ACROSS DATASETS

Table 6 presents a performance comparison between APPT, the top-performing baseline from the previous section, and our model. Both were trained on the small dataset and tested on the large dataset. The first column lists the model names, while the subsequent columns report the evaluation metrics: accuracy, precision, recall, F1, and AUC, respectively. Unlike the previous section, where APPT's metrics were taken from its original paper, the values presented here for APPT were obtained using their original codebase. We trained the APPT model on the small dataset and tested it on the large dataset to ensure a fair comparison, employing the same hyperparameters as reported in their study.

| | Accuracy | Precision | Recall | F1 | AUC |
|---|---|---|---|---|---|
| **APPT** | 60.5 | 59.1 | **90.2** | 71.4 | 55.4 |
| **Ours** | **72.7** | **72.4** | 78.7 | **75.4** | **72.3** |

*Table 6 Comparison of performance when training on the small dataset and testing on the large dataset between APPT and our model*

Our model demonstrates superior performance in this evaluation. Notably, it achieves an accuracy of 72.7%, significantly outperforming APPT (60.5%) by 12.2 percentage points. While APPT achieves a higher recall (90.2%), this comes at the cost of a significantly lower precision (59.1%), which penalizes its F1-score. Our model, however, provides a much better balance between precision (72.4%) and recall (78.7%), resulting in a superior F1-score of 75.4%, exceeding APPT's (71.4%) by 4.0 percentage points. Notably, for AUC, our model achieves 72.3%, substantially surpassing APPT's 55.4% by 16.9 points.

## 4.4 REASONING EXAMPLES

Beyond the quantitative metrics presented, one of RePaCA's key contributions is its ability to generate a transparent justification for its decisions. To illustrate this qualitative capability, this section presents two concrete examples that showcase the model's reasoning process. The outputs for both an overfitting and a correct patch are analyzed, paying special attention to the <think> block, where the model details its step-by-step analysis to reach its final answer.

### 4.4.1 Example of an Overfitting Patch

Regarding Code 1, the original method accumulates two sums (*yonscale2* and *ydotonscale2*) with the squares of ratios and then calculates a step *h* according to the ratio between the two sums or sets a minimum value if either is too small. The patch adds (on lines 5 and 6) an *if (scale.length != this.maxstep)* before updating *ydotonscale2*.

```
1      ratio = y0[j] / scale[j];
2      yonscale2 += ratio * ratio;
3      ratio = ydot0[j] / scale[j];
4    - ydotonscale2 += ratio * ratio;
5    + if ((scale.length) !=
       (org.apache.commons.math.ode.nonstiff.AdaptiveStepsizeIntegrator.thi
       s.maxstep)) {
6    +     ydotonscale2 += ratio * ratio;}
7      }
8      double h = ((yonscale2 < 1.0e-10) || (ydotonscale2 < 1.0e-10)) ?
9          1.0e-6 :
10         (0.01 * FastMath.sqrt(yonscale2 / ydotonscale2));
```

*Code 1: Code snippet of an overfitting patch example*

The model's reasoning, presented in Output 1, accurately describes the original functionality and the conditional addition, and correctly points out that the patch does not mitigate division by zero when *scale[j] == 0* (the true source of the bug). It also correctly identifies that the condition is highly specific. This level of specificity is a key indicator of overfitting, as such conditions often tailor a patch to pass a specific failing test case rather than addressing the general root cause of the bug. However, the model is incorrect in stating that *ydotonscale2* is a new variable, claiming that "The patch also introduces a new variable `ydotonscale2` that is not used anywhere in the original code, which is not a good practice", despite having previously treated this variable as already declared on line 2.

*"<think>\nThe original code calculates the square of the ratio of `y0[j]` and `scale[j]` and adds it to `yonscale2`. It then does the same for `ydot0[j]` and adds it to `ydotonscale2`. After that, it checks if either `yonscale2` or `ydotonscale2` is less than `1.0e-10`. If so, it sets `h` to `1.0e-6`. Otherwise, it calculates `h` using a formula involving the square root of the ratio of `yonscale2` to `ydotonscale2`.\n\nThe proposed patch introduces a new line `if ( ( scale . length ) ! = ( org . apache . commons . math . ode . nonstiff . adaptivestepsizeintegrator . this . maxstep ) ) { ydotonscale2 + = ratio * ratio ; }` after the second line of the original code. This line adds the square of the ratio to `ydotonscale2` if the length of `scale` is not equal to the maximum step size of the integrator. However, this addition does not address the root cause of the bug in the original code, which is the division by zero error when `scale[j]` is zero.\n\nThe patch only works for specific test cases where `scale[j]` is not zero. It does not handle the case where `scale[j]` is zero, which is the root cause of the bug in the original code. The patch also introduces a new variable `ydotonscale2` that is not used anywhere in the original code, which is not a good practice.\n\nThe patch is not generalizable and only works for specific input values of `y0[j]`, `ydot0[j]`, `scale[j]`, and the maximum step size of the integrator. The original code is more robust and handles all possible input values without introducing new bugs or side effects. \n</think>\n <answer>overfitting</answer>"*

*Output 1: Output of the model for the given Code 1 example*

### 4.4.2 Example of a Correct Patch

Code 2 refers to the *add(Complex rhs)* method of the *Complex* class, which first calls *checkNotNull* and then creates a new complex number by adding the real and imaginary parts. The patch inserts a block on lines 3 to 5 that immediately returns *NaN* if either the current object or *rhs* is *NaN*.

```
1     public Complex add(Complex rhs) throws NullArgumentException {
2         MathUtils.checkNotNull(rhs);
3  +      if ((isNaN() || rhs.isNaN()) == true) {
4  +          return NaN;
5  +      }
6         return   createComplex(real   +   rhs.getReal(),   imaginary   +
      rhs.getImaginary());
7     }
```

*Code 2: Code snippet of a correct patch example*

As shown in Output 2, the model accurately describes the original problem—the lack of NaN handling— and explains how the patch solves it in a generic manner (without depending on specific values), thereby guaranteeing a consistent result. Consequently, the model correctly recognizes and analyzes the patch, emphasizing the rationale and the causes of the change.

*"<think>\nThe original code does not handle the case where either the real or imaginary part of the complex number `rhs` is NaN (Not-a-Number). This can lead to undefined behavior or incorrect results. The patch introduces a check for NaN values in both the real and imaginary parts of `rhs`. If either part is NaN, it returns NaN. This addresses the root cause by ensuring that the addition operation does not result in undefined behavior. The patch is general and not specific to any particular input values of `a` and `b`, other than handling the NaN case.\n</think>\n<answer>correct</answer>"*

Output 2: Output of the model for the given Code 2 example

## 4.5 Ablation Studies

To validate the architectural and hyperparameter choices made during the development of RePaCA, we conducted two ablation studies. These experiments isolate particular components —namely, the base model selection and the sampling temperature— to understand their individual impact on the overall performance.

### 4.5.1 Impact of Base Model Selection

We evaluated four different lightweight LLMs from leading open-source providers to determine the most suitable backbone for the APCA task prior to GRPO fine-tuning. We selected models with parameter counts around 3B to maintain computational efficiency while ensuring sufficient reasoning capabilities. The candidates were Qwen2.5-3B (General), Qwen2.5-Coder-3B, Llama-3.2-3B and Gemma-2-2B. All were instruct versions, as the query is treated as a chat message.

Table 7 shows the baseline performance of these models on the small dataset. While the general *Qwen2.5-3B* achieved the highest raw accuracy (56.7%), it demonstrated a very low F1-score (14.4%), indicating a strong bias towards the majority class and a failure to correctly identify overfitting patches. In contrast, Qwen2.5-Coder provided the best balance, achieving the highest F1-score (38.4%) and AUC (53.6%) among the candidates.

This superiority is likely due to the "Coder" variant's pre-training on a code-heavy corpus [28], which provides a stronger initialization for understanding syntactic nuances in patch assessment. Consequently, Qwen2.5-Coder-3B was selected as the base model for the GRPO fine-tuning pipeline.

| Model | Parameters | Accuracy | F1-Score | AUC |
| --- | --- | --- | --- | --- |
| **Gemma-2** | 2B | 55.8 | 18.1 | 52.0 |
| **Llama-3.2** | 3B | 54.2 | 16.0 | 50.2 |
| **Qwen2.5** | 3B | **56.7** | 14.4 | 52.5 |
| **Qwen2.5-Coder** | 3B | 55.9 | **38.4** | **53.6** |

Table 7: Raw performance comparison of base models before GRPO fine-tuning

### 4.5.2 Impact of Temperature

The generation temperature is a critical hyperparameter in reasoning LLMs, as it controls the diversity of the Chain of Thought. We evaluated the model's performance across a temperature range from 0.0 to 1.2.

As illustrated in Figure 4, the model's performance is highly sensitive to this parameter. We observed a clear performance peak at a temperature of 0.6, where the model achieved its maximum accuracy (83.1%), F1-score (84.8%), and AUC (82.8%). These findings indicate that low temperatures cause the model to lack diversity, negatively impacting performance. Conversely, at high temperatures, the reasoning becomes incoherent and hallucinatory, leading to a sharp decline in all metrics.

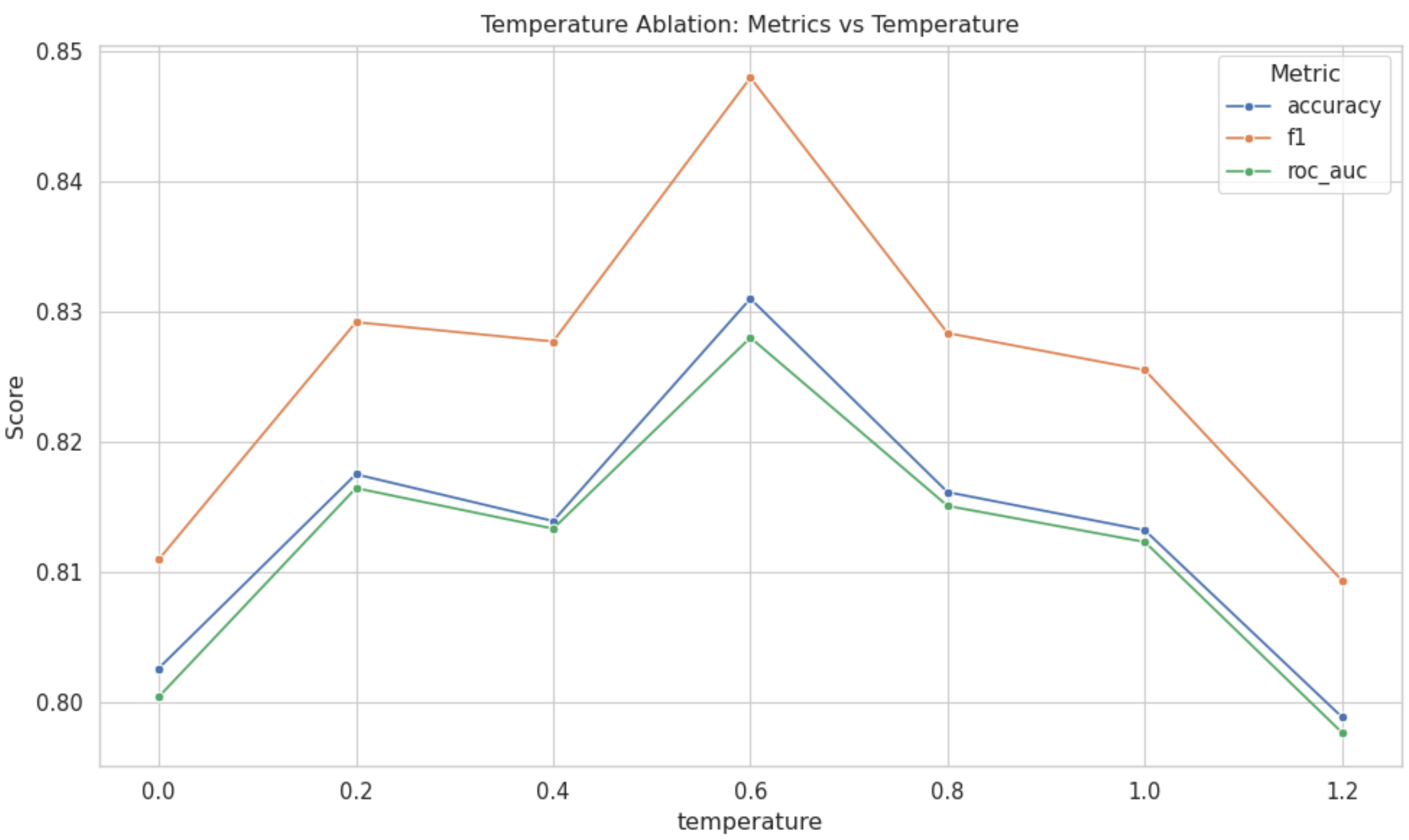

*Figure 4 Impact of sampling temperature on Accuracy, F1, and AUC metrics.*

In summary, the experimental evaluation confirms that RePaCA establishes a new state-of-the-art for static APCA, outperforming leading baselines such as APPT and INVALIDATOR across key metrics. The model also exhibits strong robustness to domain shift, while ablation studies validate the critical importance of the selected backbone and the optimized generation temperature for generating coherent reasoning traces.

# 5 Discussion

This section discusses the main contributions and limitations of our proposed approach to APCA. It highlights both quantitative improvements in performance and qualitative support for inspecting model predictions. Additionally, it addresses potential threats to validity, including reasoning failures and computational trade-offs, providing a balanced perspective on the strengths and constraints of the method.

## 5.1 Contributions

In this work, we present two primary dimensions of contribution: quantitative performance improvements and qualitative advancements in model transparency. Our study focuses on the static APCA setting, where reasoning-oriented LLM training is tailored to assess patch correctness from buggy and patched code snippets alone. Together, the performance results and generated rationales showcase not only the strength of our approach in measurable terms across established benchmarks but also its practical value for inspecting patch assessments.

From a quantitative perspective, our work has demonstrated the best overall performance on the small dataset benchmark among the techniques presented in Section 4.1. Our reasoning-capable LLM not only surpassed established static APCA techniques, such as ODS [12], Quatrain [13], and CACHE [9], but also notably outperformed the previous leading method, APPT [11], across all reported metrics (accuracy, precision, recall, F1-score, and AUC). This superior performance is largely attributed to the LLM's capacity for deeper and more semantically aware analysis of code changes, moving beyond the syntactic or structural pattern matching that characterizes many existing static methods. The GRPO fine-tuning process proved to be effective and critical for the APCA task, enabling the model to better discern the crucial differences between a correct patch and an overfitted patch.

Beyond the accuracy metrics, the robust balance achieved between precision (84.0%) and recall (85.7%) is particularly important for the practical utility of an APCA tool. High precision minimizes the risk of incorrectly discarding a correct patch, which could harm developer productivity and trust in APR systems. At the same time, high Recall ensures that a significant proportion of problematic overfitting patches, which could introduce new bugs or mask existing ones, are effectively identified. The balanced performance across these metrics, reflected in a high F1-score (84.8%), suggests that our approach provides a well-rounded and reliable assessment, managing the trade-off between falsely identifying correct patches (false positives) and failing to detect genuinely overfitting ones (false negatives). This equilibrium is vital for deploying APCA tools in real-world software development workflows [29].

It is worth noting that recent approaches based on in-context learning, such as LLM4PatchCorrect [14], are excluded from the comparison in Table 4 due to fundamental differences in their practical application scenarios. The effectiveness of that model heavily relies on retrieving semantically similar labeled patches to guide the LLM inference. Assuming the prior availability of a pool of manually checked, closely related patches for a specific bug is a theoretical assumption. This methodological premise diverges from the static approach of this work, where tools must address new or unseen bugs using only the patch itself without a direct history of similar solutions to leverage for context.

Furthermore, when evaluated using the data from INVALIDATOR [10] (as detailed in Section 4.2), our model demonstrates a significant improvement in overall metrics, achieving an accuracy of 87.8% and an F1-score of 92.6%, compared to INVALIDATOR's 81.2% and 87.0%, respectively. This gain is primarily driven by a substantial increase in recall (97.2% versus 78.9%), indicating that RePaCA is highly effective at identifying the vast majority of overfitting patches. However, this improvement comes at the cost of lower precision (88.3% versus INVALIDATOR's 96.6%). In a practical context, this means RePaCA exhibits a slightly higher tendency to falsely classify correct patches as overfitting. While this trade-off may result in the occasional discarding of a valid fix, the significant boost in recall provides a much stronger safeguard against the silent integration of overfitting patches, ultimately prioritizing the behavioral integrity of the repaired software over the raw patch acceptance rate. This benchmark was conducted using the original INVALIDATOR dataset rather than the small dataset, in order to maintain rigor and ensure consistency with their original work.

Our cross-dataset evaluation in Section 4.3 demonstrates RePaCA's robustness against domain shift. When transitioning from a controlled training environment (the small dataset, sourced from specific APR tools) to a highly heterogeneous evaluation setting (the large dataset, comprising patches from diverse open-source GitHub projects), our model significantly outperformed the baseline. While existing techniques such as APPT suffered severe performance degradation due to this distribution gap —yielding a near-random AUC of 55.4% that underscores its architectural vulnerability to out-of-distribution data— RePaCA maintained a balanced predictive capacity, achieving a 75.4% F1-score and a 72.3% AUC. This indicates that leveraging reasoning capabilities allows the model to extract a deeper semantic understanding of patch correctness, facilitating effective out-of-distribution generalization in realistic deployment scenarios where curated APR data is scarce.

Beyond its empirical performance, RePaCA offers concrete practical implications for real-world software maintenance by functioning as an accessible post-generation filter within Automated Program Repair (APR) pipelines. Its lightweight 3B-parameter architecture allows for efficient inference on standard consumer hardware, facilitating its integration into everyday development environments without prohibitive computational costs. Furthermore, its demonstrated robustness to domain shifts indicates that it can reliably assess heterogeneous patches from diverse repositories, while its explicit CoT rationales can provide developers with transparent diagnostic traces that actively support manual code audits and foster trust in the automated repair process.

Qualitatively, our methodology provides human-readable reasoning traces through the model's explicit generation of a CoT process, encapsulated within the <think>...</think> tags, as demonstrated in Section 4.4. By detailing its step-by-step analysis of the syntactic and semantic discrepancies between the buggy and fixed code, the model offers a human-readable account that can support the inspection of the tool's assessments and help analyze reasoning failures.

## 5.2 Threats to Validity

This section reflects on the practical and technical limitations observed in our approach, focusing on the nature of reasoning errors, the structural resilience of our CoT format, and the computational costs associated with our model. These insights are crucial to contextualize our contributions and outline the boundaries of applicability for real-world deployment.

A major threat to validity lies in fundamental flaws within the model's reasoning process when analyzing code patches. As shown in Code 1, we observed that sometimes the model hallucinates, inventing problems or solutions with no basis in the provided code. In other cases, it fails to correctly identify the actual change introduced; for example, the model occasionally fails to accurately perceive the diff between code snippets, rendering its subsequent analysis of correctness irrelevant or incorrect. Additionally, some information might not be provided in the code snippet's context, which leads the model to assume a specific behavior of a method or a variable throughout the program. Other critical weaknesses in reasoning include an overemphasis on superficial or irrelevant changes, which distracts from the root cause of the patch. Furthermore, faulty or even contradictory logical conclusions were observed; for instance, the reasoning may describe a patch as solving the bug (suggesting it is correct), yet the final label is assigned as "overfitting". For these reasons, RePaCA's explanations should be used to support developer inspection rather than replace it entirely.

A methodological threat in our cross-dataset evaluation is the substantial domain shift between the training and testing sets. While the small dataset contains patches generated by specific APR tools, the large dataset comprises heterogeneous patches sourced from diverse GitHub repositories. We intentionally adopted this asymmetric setup to simulate a realistic transfer scenario where high-quality curated data is scarce. Consequently, while this distribution gap inherently lowers overall metrics, it serves as a crucial benchmark for evaluating the model's architectural robustness against the out-of-distribution patches typical of real-world software maintenance.

Another internal threat to validity pertains to the reproducibility of the baseline model (APPT) within our specific experimental setup. Although we utilized the official codebase and identical hyperparameter configurations, our local reproduction of APPT —which was strictly used for the paired statistical tests and the cross-dataset evaluation— yielded slightly lower metrics than its originally published aggregate results. Such discrepancies are extensively documented in the deep learning literature and often stem from unrecorded environmental variations, including differing hardware architectures, software framework versions, or non-deterministic GPU operations [30], [31]. To ensure methodological rigor, while our broad baseline comparison relies on the externally reported metrics, our cross-dataset transfer and statistical significance analyses strictly contrast RePaCA against this local, identically-conditioned reproduction of APPT.

As explained in Section 3.2, despite the reward-based training designed to ensure correctly formatted outputs, we observed that in approximately 1% of cases during testing, the model failed to perfectly adhere to these tagging conventions (e.g., by omitting one of the tags). However, this is a relatively minor issue and can be readily addressed through simple post-processing or by implementing a re-sampling strategy for malformed outputs.

Nevertheless, the use of GRPO for fine-tuning, while demonstrably effective, inherently incurs higher computational overhead during the training phase compared to traditional APCA methods. This is primarily due to the large volume of generations required by the policy, which demands access to significant GPU resources and longer training durations. However, it is pertinent to note that the inference phase, employing our selected ~3B-parameter model, remains relatively fast and efficient. This distinction is crucial, as training is a one-time or infrequent cost, whereas inference speed is paramount for practical deployment in an APR pipeline.

To summarize, our approach provides substantial quantitative gains and qualitative transparency for static patch assessment. Although limitations exist regarding occasional reasoning hallucinations and computational training costs, these findings lay a strong foundation for future research focused on dataset enrichment and integration into real-world APR pipelines.

# 6 FUTURE WORKS

To address the identified limitations, a primary direction for improvement lies in expanding and enriching the training dataset. The current dataset is small and limited to correct and overfitting patches, which restricts the model's learning potential. Increasing its size with patches generated by a broader range of APR tools, while maintaining the quality of the original dataset, could significantly enhance performance. Equally important is the quality and depth of each sample. For example, as seen in Code 3, some patches are too small or context-poor, making it nearly impossible for the model to classify them reliably. Incorporating additional repository context —such as surrounding methods or classes— alongside complementary data representations, such as Abstract Syntax Trees (as used in ODS [12] and CACHE [9]), execution traces, or natural language explanations of the bug and intended fix, could provide the model with crucial semantic and structural context. While such hybrid datasets may improve reasoning and output quality, they could also increase the complexity and overhead of model deployment.

```
1     case ‘/’:
2  -      out.write(‘\\\\’);
3  +      if (escapesinglequote) {
4  +          out.write(‘\\\\’);
5  +      }
6         out.write(‘\\\\’);
```

*Code 3: Example of an indistinguishable overfitting patch*

Another critical step forward is to evaluate the model within an APR pipeline. Integrating RePaCA after the patch generation phase would test its utility as a filter for overfitting patches before they are presented to developers.

Furthermore, enhancing the quality of the model’s reasoning and justifications is essential. Currently, training focuses on the accuracy of the final answers and the output format, but it does not assess whether the justification process itself is logically sound. This gap in evaluation makes it difficult to fully interpret the model’s internal logic and can lead to inconsistencies. Therefore, incorporating a reward that evaluates the quality of the reasoning —validating the entire response rather than only the final result— could be beneficial. Such an approach could improve both the quality of the generated rationales and the overall performance by encouraging more coherent reasoning. In parallel, while the pretrained Qwen2.5-Coder-3B model has shown remarkable performance, evaluating more powerful models with superior explanatory and reasoning capabilities could yield more consistent CoTs. Finally, exploring models specifically trained for reasoning tasks may lead to further improvements in the observed performance.

# 7 CONCLUSIONS

In this study, we evaluated a reasoning-based APCA technique designed to mitigate patch overfitting in APR tools. Our approach leverages the reasoning capabilities of LLMs to tackle this problem, prompting them only with the necessary instructions and the buggy and fixed code snippets. Our method guides the LLM to generate a comprehensive analysis of the code differences and reason about the patch’s correctness before making a binary classification. To specifically enhance this process for the APCA task, we fine-tuned the LLM using the GRPO algorithm.

In our evaluation on the small dataset, we obtained the best performance among the static APCA baselines considered, achieving an accuracy of 83.1% and an F1-score of 84.8%, and significantly surpassing previous leading static techniques in this setting. The model also demonstrates greater robustness to domain shift compared to APPT when training and testing on different datasets. Beyond its quantitative performance, RePaCA provides human-readable reasoning traces that facilitate the inspection of each assessment. These traces should be treated as auxiliary explanations rather than definitive evidence of the model's internal decision process.

In practical terms, RePaCA can serve as a post-generation filter in APR workflows, reducing the need for manual inspection and mitigating the risk of accepting overfitting patches. Based on these results, future work should emphasize scaling the dataset in both quality and quantity, incorporating richer contextual information for context-poor patches and improving the supervision of the model's reasoning process during training.

**Data availability statement**

The datasets are not publicly available due to ongoing research use, but are available from the corresponding author upon reasonable request.

**Acknowledgements**

This work was co-supported both by the project "Reparación automática de código fuente mediante modelos generativos de Procesamiento de Lenguaje Natural" (SBPLY/23/180225/000063) through the call "Ayudas para la realización de proyectos de investigación científica y transferencia tecnológica 2023"; by the project "Tecnologías Inteligentes para la Fabricación, el diseño y las Operaciones en entornos iNdustriales" (TIFON, PLEC2023-010251) through the call "Proyectos de I+D+i en líneas estratégicas - Transmisiones 2023"; and by the project "Detección de Errores y Vulnerabilidades en el Software con Autoencoders Dispersos" (PIUAH25/IA-026) through the call "Ayudas para la realización de proyectos de investigación UAH".

## Author biography

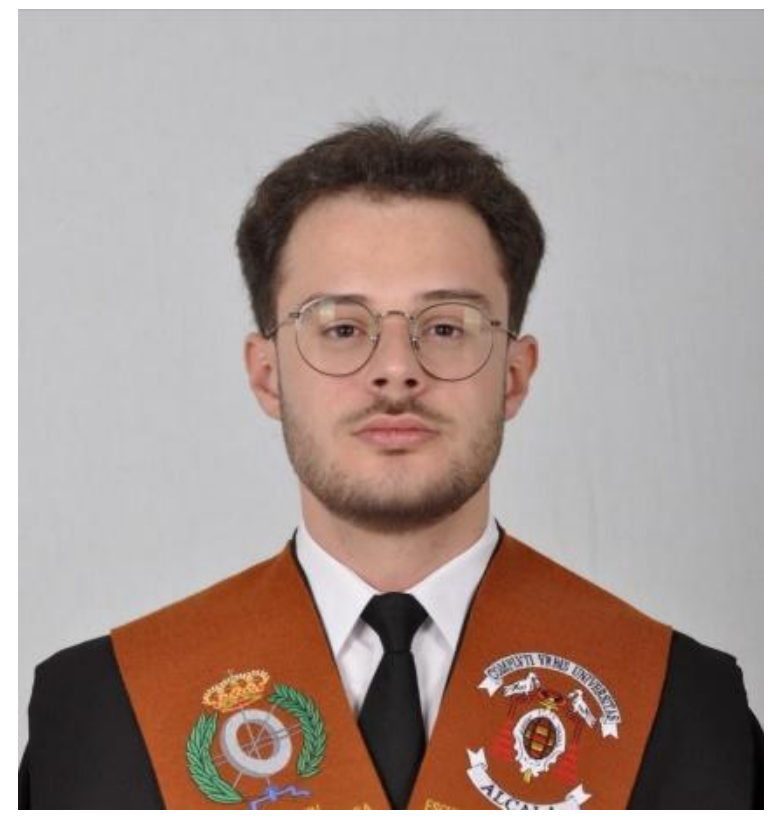

**Marcos Fuster-Pena** is currently completing his B.Sc. degree in Computer Science at the University of Alcalá, graduating in 2026. He will soon pursue further academic specialization in Natural Language Processing (NLP). He is currently part of the INTELIA (Interaction Technologies and Artificial Intelligence) Research Lab. His primary research interests lie in the intersection of Artificial Intelligence and Software Engineering, focusing specifically on Automated Program Repair through the development of Agent systems and reinforcement learning methods.

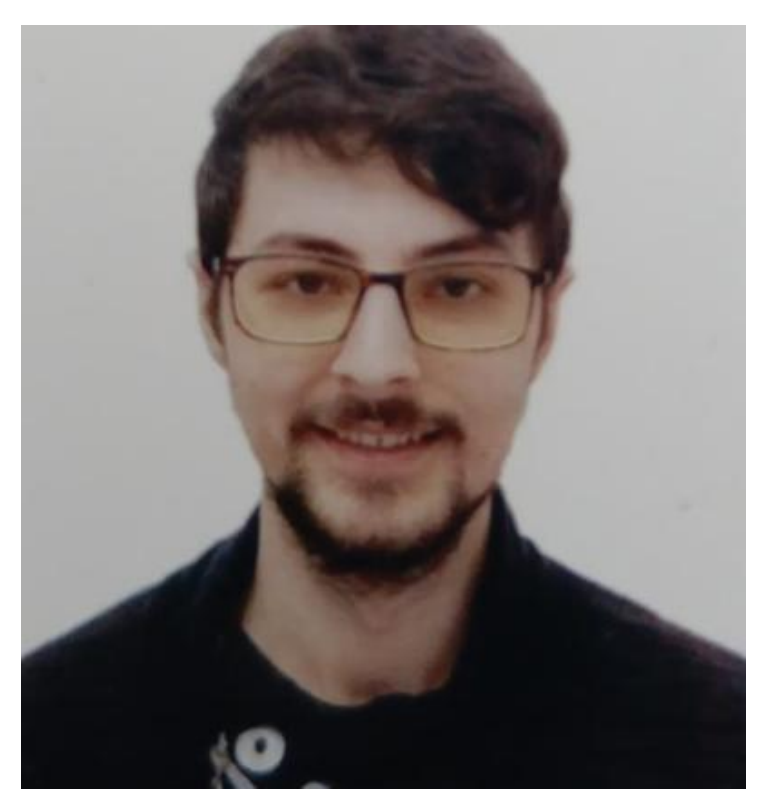

**David de-Fitero-Dominguez** received the B.Sc. and M.Sc. degrees in computer science from the University of Alcala, Spain, in 2022 and 2023, respectively, where he is currently pursuing the Ph.D. degree. He was awarded with the ‘‘Best Graduate Student of the Year’’ mention in the B.Sc. degree. He has participated in several journals and conference publications and also a Researcher in different research projects. Currently, he is part of the Intelia Research Team, delving into the areas of artificial intelligence, natural language processing, and large language models.

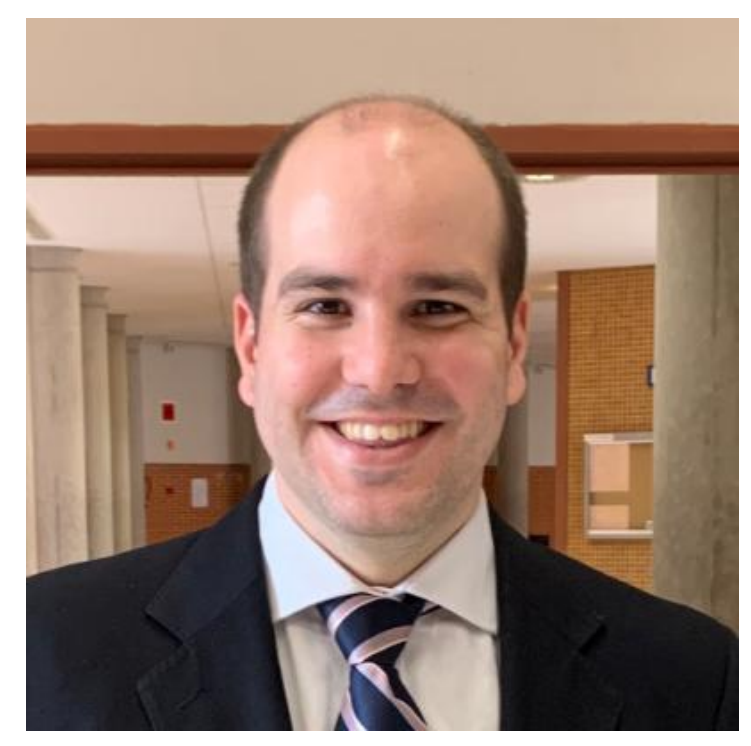

**Antonio Garcia-Cabot** is a Full Professor in the Computer Science Department at the University of Alcala (Spain). He is the Director of the INTELIA research lab, which is classified as a high-performance group. His research interests include Natural Language Processing, Artificial Intelligence, Educational Technology and User Experience. He has published more than 100 conference papers and articles in high-impact journals (39 of them published in ISI JCR journals). He has participated as a researcher in more than 34 national and international projects, in 9 of them as a principal investigator (1 of them funded by the European Commission and 3 national research projects). He has also participated in 52 contracts and projects with other companies or institutions, 18 of them as a principal researcher. He has supervised 4 doctoral theses: with special mention, 1 joint supervised with UFM University, 1 with international PhD mention and 1 at the University of Zagreb (Croatia). He has supervised more than 80 final undergraduate and graduate theses. He was a Visiting Scholar in the Comparative Media Studies / Writing Department at the Massachusetts Institute of Technology (USA) in 2024. He stayed during his PhD studies in Lund (Sweden), Oxford (United Kingdom) and Helsinki (Finland) for a total of 9 months; and he has performed three postdoctoral stays (in Helsinki and Lisbon) for a total of 13 months in total.

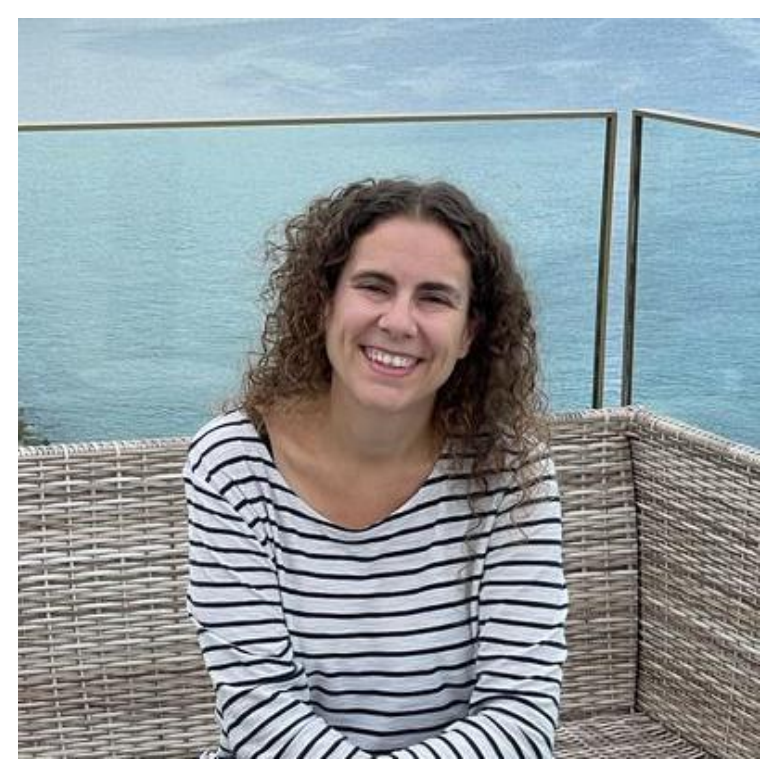

**Eva Garcia-Lopez** is an Associate Professor in the Department of Computer Science at the University of Alcala (Spain). She is the co-director of the INTELIA research lab. She has been a Visiting Scholar in the Comparative Media Studies / Writing Department at the Massachusetts Institute of Technology (USA). Her research interests include artificial intelligence, natural language processing, educational technology and user experience. She has published more than 100 conference papers and articles in high-impact journals, including 36 papers published in ISI JCR-indexed journals, 24 papers in CORE-ranked conferences, and 2 intellectual property records. She has participated in 26 publicly funded research projects (in 5 of them as a principal investigator), including 8 funded by the European Commission; and more than 40 contracts with companies and private institutions, being the principal investigator in three of them. During her PhD studies, she spent a total of 12 months at several universities abroad, in Lund (Sweden), Oxford (United Kingdom) and Helsinki (Finland); and she performed two postdoctoral stays (in Helsinki and Lisbon), for a total of 12 months in total.